\documentclass[conference]{IEEEtran} 
\usepackage{graphicx}
\usepackage{epsfig}
\usepackage{epstopdf}
\usepackage{amsmath,amsthm,amsfonts,amssymb}
\usepackage{multirow}
\usepackage{cite}
\usepackage{verbatim}
\usepackage{tikz}
\usetikzlibrary{arrows}
\usetikzlibrary{backgrounds}
\usepackage{ellipsis}
\usepackage{xcolor}
\usepackage{xifthen}
\usepackage{cancel}
\usepackage{subfig}
\usetikzlibrary{shapes,arrows,intersections}
\usetikzlibrary{patterns,decorations}
\usetikzlibrary{mindmap}
\usetikzlibrary{calc}
\usepackage{ellipsis}
\usetikzlibrary{decorations.pathreplacing,decorations.markings,shapes.geometric,arrows,decorations.pathmorphing,decorations.footprints,fadings,calc,trees,mindmap,shadows,decorations.text,patterns,positioning,shapes,matrix,fit,chains}

\theoremstyle{definition} 
\theoremstyle{definition} 
\theoremstyle{definition} 
\theoremstyle{definition} 

\newcommand{\set}[1]{\mathcal{#1}}
\newcommand{\Fb}{\mathbb{F}_2}
\newcommand{\CRC}{\mathsf{CRC}}

\DeclareMathOperator{\sgn}{sgn}

\tikzset
{
	BlocksStyle/.style =
	{
		shape			= rectangle,			% shape
		rounded corners	= 0.0cm,				% radius of the rounded corner
		minimum height	= 0.7cm,				% | minimum size of the node
		minimum width	= 0.9cm,				% |
		rotate			= 0,					% angle of rotation
		scale			= 1.0,					% scaling factor
		draw			= black,				% draw the border with this color
		line width		= 0.00cm,				% thickness
		align			= center,				% text alignment
		text			= black,				% color of the fonts
		font			= \normalsize\normalfont,	% shape and dimension of the font
		inner xsep		= 0.2cm,				% min. dist. btw text and borders along x dimension
		inner ysep		= 0.2cm,				% min. dist. btw text and borders along x dimension
	}
}

\tikzset
{
	BlocksStyleb/.style =
	{
		shape			= rectangle,			% shape
		rounded corners	= 0.0cm,				% radius of the rounded corner
		minimum height	= 0.7cm,				% | minimum size of the node
		minimum width	= 0.9cm,				% |
		rotate			= 0,					% angle of rotation
		scale			= 1.0,					% scaling factor
		draw			= black,				% draw the border with this color
		line width		= 0.02cm,				% thickness
		align			= center,				% text alignment
		text			= black,				% color of the fonts
		font			= \normalsize\normalfont,	% shape and dimension of the font
		inner xsep		= 0.2cm,				% min. dist. btw text and borders along x dimension
		inner ysep		= 0.2cm,				% min. dist. btw text and borders along x dimension
	}
}

\tikzset
{
	WideBlocksStyle/.style =
	{
		BlocksStyle,
		text width		= 2.0cm,				% max. width of the text
	}
}
\pgfmathsetseed{1}
\usetikzlibrary{shapes,positioning,arrows,calc,graphs}
\usetikzlibrary{decorations.pathreplacing,decorations.markings,shapes.geometric}
\tikzset{naming/.style={align=center,font=\small}}
\tikzset{antenna/.style={insert path={-- coordinate (ant#1) ++(0,0.25) -- +(135:0.25) + (0,0) -- +(45:0.25)}}}
\tikzset{station/.style={naming,draw,shape=dart,shape border rotate=90, minimum width=10mm, minimum height=10mm,outer sep=0pt,inner sep=3pt}}
\tikzset{mobile/.style={naming,draw,shape=rectangle,minimum width=12mm,minimum height=6mm, outer sep=0pt,inner sep=3pt}}
\tikzset{radiation/.style={{decorate,decoration={expanding waves,angle=90,segment length=4pt}}}}

\tikzset
{
	HighlightingStyle/.style =
	{
		color				= blue,	
		line width			= 0.04cm,			% thickness
		arrows				= -,				% starting-ending arrows
		dotted,
	}
}	% no semicolons are here required

\tikzset
{
	HighlightingStyleD/.style =
	{
		color				= blue,	% color
		line width			= 0.04cm,			% thickness
		arrows				= -,				% starting-ending arrows
		dashed,
	}
}	% no semicolons are here required

\tikzset
{
	HighlightingStyleB/.style =
	{
		color				= black,	% color
		line width			= 0.04cm,			% thickness
		arrows				= -,				% starting-ending arrows
		solid,
	}
}

\tikzset
{
	HighlightingStyleE/.style =
	{
		color				= red,	% color
		line width			= 0.05cm,			% thickness
		arrows				= -,				% starting-ending arrows
		dashed,
	}
}	
\tikzset
{
	HighlightingStyleC/.style =
	{
		color				= black,
		fill=white,
		text=black,	% color
		line width			= 0.05cm,			% thickness
		arrows				= -,				% starting-ending arrows
 		rounded corners		= 0.0cm,			%
		solid,
	}
}

\tikzset
{
	LinesStyle/.style =
	{
		color				= black,	% color
		line width			= 0.02cm,			% thickness
		solid,
	}
}	

\tikzset
{
	LinesStyleC/.style =
	{
		color				= black,	% color
		line width			= 0.08cm,			% thickness
		solid,
	}
}

\tikzset
{
	LinesStyleR/.style =
	{
		color				= black,	% color
		line width			= 0.08cm,			% thickness
		densely dotted,
	}
}

\tikzset
{
	LinesStyleE/.style =
	{
		color				= red,	% color
		line width			= 0.06cm,			% thickness
		solid,
	}
}

\tikzset
{
	LinesStyleb/.style =
	{
		color				= black,	% color
		line width			= 0.02cm,			% thickness
		arrows				= -latex',			% starting-ending arrows
 		shorten <			= 0.1cm,			% shorten the ending point
		solid,
	}
}

\tikzset
{
	SumNodesStyle/.style =
	{
		shape			= circle,				% shape
		minimum size	= 0.1cm,				% | minimum size of the node
		rotate			= 0,					% angle of rotation
		scale			= 0.6,					% scaling factor
		draw			= black,				% draw the border with this color
		line width		= 0.02cm,				% thickness
		text			= black,				% color of the fonts
		font			= \normalsize\normalfont,	% shape and dimension of the font
	}
}

\begin{document}

\title{Low-Complexity Puncturing and \\ Shortening of Polar Codes}

\author{\IEEEauthorblockN{Valerio Bioglio, Fr\'ed\'eric Gabry, Ingmar Land}
\IEEEauthorblockA{Mathematical and Algorithmic Sciences Lab\\ France Research Center, Huawei Technologies Co. Ltd.\\
Email: $\{$valerio.bioglio,frederic.gabry,ingmar.land$\}$@huawei.com}} 

\maketitle

\begin{abstract}
In this work, we address the low-complexity construction of shortened and punctured polar codes from a unified view.
While several independent puncturing and shortening designs were attempted in the literature, our goal is a unique, low-complexity construction encompassing both techniques in order to achieve any code length and rate. 
We observe that our solution significantly reduces the construction complexity as compared to state-of-the-art solutions while providing a block error rate performance comparable to constructions that are highly optimized for specific lengths and rates. 
This makes the constructed polar codes highly suitable for practical application in future communication systems requiring a large set of polar codes with different lengths and rates.
\end{abstract}

\section{Introduction}
\label{sec:intro}
Polar codes were introduced by Arikan in his seminal work \cite{polar} as a new class of channel codes that are capacity-achieving over various classes of channels. 
Their encoding and decoding complexity is of order $O(N \log N)$ in terms of the code length $N$, which constituted a breakthrough in coding theory.
The state-of-the-art polar decoder uses successive cancellation (SC) decoding \cite{polar} in connection with list (SCL) decoding \cite{list_decoding} and possibly an additional cyclic redundancy check (CRC) to improve the distance properties of the code and hence improve the list decoding performance, referred to as CRC-aided SCL decoding \cite{CRC_aid}.

In their original construction, polar codes only allowed code lengths that are powers of two, while the code dimension (the number of information bits) is arbitrary. 
This restriction on the code lengths is a major drawback of polar codes, which needs to be overcome for practical applications. 
Recent works have proposed ways to modify polar codes in order to achieve any code length, using the techniques of puncturing and shortening from coding theory.
In puncturing, one or more code bits are not transmitted. 
The decoder typically handles punctured bits as erased and applies the decoder of the mother code. 
In shortening, a sub-code of the mother code is used in which one or more code bits are restricted to a fixed value, typically zero, and not transmitted. 
The decoder applies the decoder of the mother code, where the shortened code bits are assumed to be known.

Numerous practical solutions were proposed in literature to achieve any codelength with polar codes. 
One option is to design the frozen set of the mother polar code based on the bits that will be punctured or shortened. 
In particular, this method is proposed in \cite{wang_liu} for the shortening of polar codes, while in \cite{short_mat,chen_kai_punc} the same method is chosen for the puncturing of polar codes. 
In \cite{wang_liu}, the shortening pattern is given by the last bits of the mother polar codes, while in \cite{short_mat,chen_kai_punc} the puncturing pattern is generated according to a certain heuristic. 
In both cases, a density evolution (DE) algorithm is run to find the optimal frozen set given the puncturing or shortening set. 
However, the frozen set needs to be calculated at every encoding and every decoding; 
this decreases the suitability of that solution for practical implementation, since the DE algorithm adds significant complexity to both the encoder and the decoder. 
Hence, even though the error-rate performance of this technique is acceptable, the high complexity and resulting latency due to the DE calculation makes it unattractive for hardware implementation. 

An alternative approach proposed in \cite{isit_punc,ARQ_punct} is to select the bits to puncture or shorten given the frozen set of the mother polar code. 
In \cite{isit_punc}, an algorithm is presented to find an optimal puncturing pattern leading to the selected frozen set. 
Similarly, in \cite{ARQ_punct}, the authors proposed to puncture bits with indices corresponding to the less reliable bits, based on the conjecture that these bits are less important than the others. 
In general, designing the puncturing set based on the frozen set allows a significant complexity reduction, albeit at the cost of an increased BLock Error Rate (BLER). 

Finally, a joint optimization technique is presented in \cite{short_milo} for the shortening of polar codes, where the frozen set and the shortening pattern are jointly optimized. 
A search over all possible shortening patterns is performed, largely reducing the search complexity by exploiting certain symmetries of the polar encoder. 
This technique yields an optimized BLER performance, but at a high cost in terms of memory and complexity requirements. 

To summarize, the main shortcomings of existing techniques is the lack of general structure in the frozen, puncturing and shortening sets obtained using these methods. 
In particular, every code length and dimension corresponds to different frozen, puncturing and shortening sets. 
These sets have to either be stored, requiring a large memory for storage, or be calculated on the fly, resulting in large computational complexity. 
Hence these methods are not practical to their large description complexity.

In this paper, we propose a low-complexity technique to modify polar codes. 
In particular, we propose to use puncturing and shortening sets based on the bit-reversal permutation to define a family of polar codes for arbitrary lengths and rates. 
This construction is obtained based on observations on the behavior of the puncturing and shortening sets, which we present in the paper. 
The description of the family of polar codes is stored in a compact way, as opposed to other solutions that require large tables or complex algorithms to generate the frozen, puncturing or shortening sets.
The low-complexity description allows for practical application in systems where a large set of polar codes with different lengths and rates is required.
The error-rate performance of the proposed polar codes is comparable to existing solutions, where the individual codes are optimized for the desired length and rate. 

This paper is organized as follows. In Section \ref{sec:polar} we provide some preliminaries on polar codes while in Section \ref{sec:punct} and Section \ref{sec:short} we give an overview of puncturing and shortening techniques. 
In Section \ref{sec:prop}, we propose our unified framework. 
In Section \ref{sec:num}, we evaluate numerically our proposed construction, and finally Section \ref{sec:conclusions} concludes this paper.

\section{Preliminaries on Polar Codes}
\label{sec:polar}
\begin{figure}[t!]
\begin{center}
\resizebox{0.52\textwidth}{!}{\begin{tikzpicture}
[
	xscale	= 1,	% to scale horizontally everything but the text
	yscale	= 1,	% to scale vertically everything but the text
]

% ------------------------------------------------------
% NODES DEFINITION
\matrix
(nMatrix)
[
	row sep		= 0.6cm,
	column sep	= 2.1cm, ampersand replacement = \&
]
{

\node (n01) {$v_1$};  \&[-4ex]\node (n02)  {};\&[-4ex]\node (n03) {};\&[-4ex]\node (n04) {};\&[-4ex]\node (n05) {};\&[-4ex]\node (n06) {};\&[-4ex]\node (n07) {};\&[-4ex]\node (n08) {$x_1$};
	\\
\node (n11) {$v_2$};\&\node(n12) {};\&\node (n13) {};\&\node (n14) {};\&\node (n15) {};\&\node (n16) {};\&\node (n17) {};\&\node (n18) {$x_2$}; \\

\node (n21) {$v_3$};\&\node(n22) {};\&\node(n23) {};\&\node (n24) {};\&\node (n25) {};\&\node (n26) {};\&\node (n27) {};\&\node (n28) {$x_3$};\\
% row 4
\node (n31) {$v_4$};\&\node(n32) {};\&\node(n33) {};\&\node(n34) {};\&\node (n35) {};\&\node (n36) {};\&\node (n37) {};\&\node (n38) {$x_4$};\\
% row 5
\node (n41) {$v_5$};\&\node(n42) {};\&\node(n43) {};\&\node(n44) {};\&\node(n45) {};\&\node (n46) {};\&\node (n47) {};\&\node (n48) {$x_5$};\&\\
% row 6
\node (n51) {$v_6$};\&\node (n52) {};\&\node(n53) {};\&\node(n54) {};\&\node(n55) {};\&\node (n56) {};\&\node (n57) {};\&\node(n58) {$x_6$};\&\\
\node (n61) {$v_7$};\&\node (n62) {};\&\node(n63) {};\&\node(n64) {};\&\node(n65) {};\&\node (n66) {};\&\node (n67) {};\&\node(n68) {$x_7$};\&\\
\node (n71) {$v_8$};\&\node (n72) {};\&\node(n73) {};\&\node(n74) {};\&\node(n75) {};\&\node (n76) {};\&\node (n77) {};\&\node(n78) {$x_8$};\&\\
};

% ------------------------------------------------------
% PATHS

%\draw[decorate,decoration={brace,amplitude=10pt},rotate=180] ($(n06)+ (0.0cm,-0.2cm)$) -- ($(n56)+ (0cm,0.2cm)$) node [pos=0.5, xshift = 0.5cm,rotate=270] {$L$ smallest metrics};

%%% -------------------------
%% auxiliary nodes

%\node [coordinate, xshift = -0.cm, yshift =  0.1cm] (nAux1) at (n02.north west) {};
%\node [coordinate, xshift =  0.cm, yshift =  -0.1cm] (nAux2) at (n153.south west) {};
%\draw [HighlightingStyle] (nAux1) -| (nAux2) -|  (nAux1)
%node [below, pos = 0.21] {Extract};

\draw [LinesStyle] (n01) -- (n08) ;

\draw [LinesStyle] (n11) -- ($(n13.west) + (-.3cm,-0cm)$) ;
\draw [LinesStyle] ($(n15.west) + (-.3cm,-0cm)$) -- ($(n13.east) + (.3cm,-0cm)$) ;
\draw [LinesStyle] ($(n17.west) + (-.3cm,-0cm)$) -- ($(n15.east) + (.3cm,-0cm)$) ;
\draw [LinesStyle] ($(n13.west) + (-.3cm,-0cm)$) --  ($(n43.east) + (.3cm,-0cm)$);
\draw [LinesStyle] ($(n15.west) + (-.3cm,-0cm)$) -- ($(n25.east) + (.3cm,-0cm)$) ;
\draw [LinesStyle] ($(n17.west) + (-.3cm,-0cm)$) -- ($(n47.east) + (.3cm,-0cm)$) ;
\draw [LinesStyle] ($(n17.east) + (.3cm,-0cm)$) -- (n18) ;

\draw [LinesStyle] (n21) -- ($(n23.west) + (-.3cm,-0cm)$) ;
\draw [LinesStyle] ($(n25.west) + (-.3cm,-0cm)$) -- ($(n23.east) + (.3cm,-0cm)$) ;
\draw [LinesStyle] ($(n27.west) + (-.3cm,-0cm)$) -- ($(n25.east) + (.3cm,-0cm)$) ;
\draw [LinesStyle] ($(n23.west) + (-.3cm,-0cm)$) --  ($(n13.east) + (.3cm,-0cm)$);
\draw [LinesStyle] ($(n25.west) + (-.3cm,-0cm)$) -- ($(n15.east) + (.3cm,-0cm)$) ;
\draw [LinesStyle] ($(n27.west) + (-.3cm,-0cm)$) -- ($(n27.east) + (.3cm,-0cm)$) ;
\draw [LinesStyle] ($(n27.east) + (.3cm,-0cm)$) -- (n28) ;

\draw [LinesStyle] (n31) -- ($(n33.west) + (-.3cm,-0cm)$) ;
\draw [LinesStyle] ($(n35.west) + (-.3cm,-0cm)$) -- ($(n33.east) + (.3cm,-0cm)$) ;
\draw [LinesStyle] ($(n37.west) + (-.3cm,-0cm)$) -- ($(n35.east) + (.3cm,-0cm)$) ;
\draw [LinesStyle] ($(n33.west) + (-.3cm,-0cm)$) --  ($(n53.east) + (.3cm,-0cm)$);
\draw [LinesStyle] ($(n35.west) + (-.3cm,-0cm)$) -- ($(n35.east) + (.3cm,-0cm)$) ;
\draw [LinesStyle] ($(n37.west) + (-.3cm,-0cm)$) -- ($(n67.east) + (.3cm,-0cm)$) ;
\draw [LinesStyle] ($(n37.east) + (.3cm,-0cm)$) -- (n38) ;

\draw [LinesStyle] (n41) -- ($(n43.west) + (-.3cm,-0cm)$) ;
\draw [LinesStyle] ($(n45.west) + (-.3cm,-0cm)$) -- ($(n43.east) + (.3cm,-0cm)$) ;
\draw [LinesStyle] ($(n47.west) + (-.3cm,-0cm)$) -- ($(n45.east) + (.3cm,-0cm)$) ;
\draw [LinesStyle] ($(n43.west) + (-.3cm,-0cm)$) --  ($(n23.east) + (.3cm,-0cm)$);
\draw [LinesStyle] ($(n45.west) + (-.3cm,-0cm)$) -- ($(n45.east) + (.3cm,-0cm)$) ;
\draw [LinesStyle] ($(n47.west) + (-.3cm,-0cm)$) -- ($(n17.east) + (.3cm,-0cm)$) ;
\draw [LinesStyle] ($(n47.east) + (.3cm,-0cm)$) -- (n48) ;

\draw [LinesStyle] (n51) -- ($(n53.west) + (-.3cm,-0cm)$) ;
\draw [LinesStyle] ($(n55.west) + (-.3cm,-0cm)$) -- ($(n53.east) + (.3cm,-0cm)$) ;
\draw [LinesStyle] ($(n57.west) + (-.3cm,-0cm)$) -- ($(n55.east) + (.3cm,-0cm)$) ;
\draw [LinesStyle] ($(n53.west) + (-.3cm,-0cm)$) --  ($(n63.east) + (.3cm,-0cm)$);
\draw [LinesStyle] ($(n55.west) + (-.3cm,-0cm)$) -- ($(n65.east) + (.3cm,-0cm)$) ;
\draw [LinesStyle] ($(n57.west) + (-.3cm,-0cm)$) -- ($(n57.east) + (.3cm,-0cm)$) ;
\draw [LinesStyle] ($(n57.east) + (.3cm,-0cm)$) -- (n58) ;

\draw [LinesStyle] (n61) -- ($(n63.west) + (-.3cm,-0cm)$) ;
\draw [LinesStyle] ($(n65.west) + (-.3cm,-0cm)$) -- ($(n63.east) + (.3cm,-0cm)$) ;
\draw [LinesStyle] ($(n67.west) + (-.3cm,-0cm)$) -- ($(n65.east) + (.3cm,-0cm)$) ;
\draw [LinesStyle] ($(n63.west) + (-.3cm,-0cm)$) --  ($(n33.east) + (.3cm,-0cm)$);
\draw [LinesStyle] ($(n65.west) + (-.3cm,-0cm)$) -- ($(n55.east) + (.3cm,-0cm)$) ;
\draw [LinesStyle] ($(n67.west) + (-.3cm,-0cm)$) -- ($(n37.east) + (.3cm,-0cm)$) ;
\draw [LinesStyle] ($(n67.east) + (.3cm,-0cm)$) -- (n68) ;

\draw [LinesStyle] (n71) -- (n78) ;

%\node [coordinate, xshift = -0.3cm, yshift =  0.3cm] (nAux0) at (n03.north west) {};
%\node [coordinate, xshift =  0.3cm, yshift =  -0.3cm] (nAux00) at (n73.south east) {};
%\draw [HighlightingStyle]  (nAux0) -| (nAux00) -|  (nAux0)
%node [below, pos = 0.21] {$P_3$};
%
%
%\node [coordinate, xshift = -0.3cm, yshift =  0.3cm] (nAux1) at (n05.north west) {};
%\node [coordinate, xshift =  0.3cm, yshift =  -0.3cm] (nAux2) at (n75.south east) {};
%\draw [HighlightingStyle]  (nAux1) -| (nAux2) -|  (nAux1)
%node [below, pos = 0.21] {$P_2$};
%
%\node [coordinate, xshift = -0.3cm, yshift =  0.3cm] (nAux3) at (n07.north west) {};
%\node [coordinate, xshift =  0.3cm, yshift =  -0.3cm] (nAux4) at (n77.south east) {};
%\draw [HighlightingStyle] (nAux3) -| (nAux4) -|  (nAux3)
%node (Pnode) [below, pos = 0.21] {$P_1$};

\node [coordinate, xshift = -0.3cm, yshift =  0.3cm] (nAux3) at (n07.north west) {};
\node [coordinate, xshift =  -0.2cm, yshift =  -0.5cm] (Pnode) at (n77.south east) {};
%\draw [HighlightingStyle,color=white] (nAux3) -| (nAux4) -|  (nAux3)
%node (Pnode) [below, pos = 0.21] {};

\node [coordinate, xshift = -0.3cm, yshift =  0.2cm] (nAux5) at (n02.north west) {};
\node [coordinate, xshift = 0.2cm, yshift =  -0.2cm] (nAux6) at (n12.south east) {};
\draw [HighlightingStyleB,fill=white] (nAux5) -| (nAux6) -|  (nAux5)
node (Pnode1) [xshift=0.4cm, yshift =-.9cm] {$T_2$};

\node [coordinate, xshift = -0.3cm, yshift =  0.2cm] (nAux7) at (n22.north west) {};
\node [coordinate, xshift = 0.2cm, yshift =  -0.2cm] (nAux8) at (n32.south east) {};
\draw [HighlightingStyleB,fill=white] (nAux7) -| (nAux8) -|  (nAux7)
node (Pnode1) [xshift=0.4cm, yshift =-.9cm] {$T_2$};

\node [coordinate, xshift = -0.3cm, yshift =  0.2cm] (nAux9) at (n42.north west) {};
\node [coordinate, xshift = 0.2cm, yshift =  -0.2cm] (nAux10) at (n52.south east) {};
\draw [HighlightingStyleB,fill=white] (nAux9) -| (nAux10) -|  (nAux9)
node (Pnode1) [xshift=0.4cm, yshift =-.9cm] {$T_2$};

\node [coordinate, xshift = -0.3cm, yshift =  0.2cm] (nAux11) at (n62.north west) {};
\node [coordinate, xshift = 0.2cm, yshift =  -0.2cm] (nAux12) at (n72.south east) {};
\draw [HighlightingStyleB,fill=white] (nAux11) -| (nAux12) -|  (nAux11)
node (Pnode1) [xshift=0.4cm, yshift =-.9cm] {$T_2$};

\node [coordinate, xshift = -0.3cm, yshift =  0.2cm] (nAux5b) at (n04.north west) {};
\node [coordinate, xshift = 0.2cm, yshift =  -0.2cm] (nAux6b) at (n14.south east) {};
\draw [HighlightingStyleB,fill=white] (nAux5b) -| (nAux6b) -|  (nAux5b)
node (Pnode1) [xshift=0.4cm, yshift =-.9cm] {$T_2$};

\node [coordinate, xshift = -0.3cm, yshift =  0.2cm] (nAux7b) at (n24.north west) {};
\node [coordinate, xshift = 0.2cm, yshift =  -0.2cm] (nAux8b) at (n34.south east) {};
\draw [HighlightingStyleB,fill=white] (nAux7b) -| (nAux8b) -|  (nAux7b)
node (Pnode1) [xshift=0.4cm, yshift =-.9cm] {$T_2$};

\node [coordinate, xshift = -0.3cm, yshift =  0.2cm] (nAux9b) at (n44.north west) {};
\node [coordinate, xshift = 0.2cm, yshift =  -0.2cm] (nAux10b) at (n54.south east) {};
\draw [HighlightingStyleB,fill=white] (nAux9b) -| (nAux10b) -|  (nAux9b)
node (Pnode1) [xshift=0.4cm, yshift =-.9cm] {$T_2$};

\node [coordinate, xshift = -0.3cm, yshift =  0.2cm] (nAux11b) at (n64.north west) {};
\node [coordinate, xshift = 0.2cm, yshift =  -0.2cm] (nAux12b) at (n74.south east) {};
\draw [HighlightingStyleB,fill=white] (nAux11b) -| (nAux12b) -|  (nAux11b)
node (Pnode1) [xshift=0.4cm, yshift =-.9cm] {$T_2$};

\node [coordinate, xshift = -0.3cm, yshift =  0.2cm] (nAux5c) at (n06.north west) {};
\node [coordinate, xshift = 0.2cm, yshift =  -0.2cm] (nAux6c) at (n16.south east) {};
\draw [HighlightingStyleB,fill=white] (nAux5c) -| (nAux6c) -|  (nAux5c)
node (Pnode1) [xshift=0.4cm, yshift =-.9cm] {$T_2$};

\node [coordinate, xshift = -0.3cm, yshift =  0.2cm] (nAux7c) at (n26.north west) {};
\node [coordinate, xshift = 0.2cm, yshift =  -0.2cm] (nAux8c) at (n36.south east) {};
\draw [HighlightingStyleB,fill=white] (nAux7c) -| (nAux8c) -|  (nAux7c)
node (Pnode1) [xshift=0.4cm, yshift =-.9cm] {$T_2$};

\node [coordinate, xshift = -0.3cm, yshift =  0.2cm] (nAux9c) at (n46.north west) {};
\node [coordinate, xshift = 0.2cm, yshift =  -0.2cm] (nAux10c) at (n56.south east) {};
\draw [HighlightingStyleB,fill=white] (nAux9c) -| (nAux10c) -|  (nAux9c)
node (Pnode1) [xshift=0.4cm, yshift =-.9cm] {$T_2$};

\node [coordinate, xshift = -0.3cm, yshift =  0.2cm] (nAux11c) at (n66.north west) {};
\node [coordinate, xshift = 0.2cm, yshift =  -0.2cm] (nAux12c) at (n76.south east) {};
\draw [HighlightingStyleB,fill=white] (nAux11c) -| (nAux12c) -|  (nAux11c)
node (Pnode1) [xshift=0.4cm, yshift =-.9cm] {$T_2$};

%\node[left of = n02,yshift=0.18cm,xshift=-0.05cm] (LLR20) {\footnotesize{LLR$(2,0)$}};
%\node[left of = n12,yshift=0.18cm,xshift=-0.05cm] (LLR21) {\footnotesize{LLR$(2,1)$}};
%\node[left of = n22,yshift=0.18cm,xshift=-0.05cm] (LLR22) {\footnotesize{LLR$(2,2)$}};
%\node[left of = n32,yshift=0.18cm,xshift=-0.05cm] (LLR23) {\footnotesize{LLR$(2,3)$}};
%\node[left of = n42,yshift=0.18cm,xshift=-0.05cm] (LLR24) {\footnotesize{LLR$(2,4)$}};
%\node[left of = n52,yshift=0.18cm,xshift=-0.05cm] (LLR25) {\footnotesize{LLR$(2,5)$}};
%
%\node[left of = n04,yshift=0.18cm,xshift=-0.05cm] (LLR10) {\footnotesize{LLR$(1,0)$}};
%\node[left of = n14,yshift=0.18cm,xshift=-0.05cm] (LLR11) {\footnotesize{LLR$(1,1)$}};
%\node[left of = n24,yshift=0.18cm,xshift=-0.05cm] (LLR12) {\footnotesize{LLR$(1,2)$}};
%\node[left of = n34,yshift=0.18cm,xshift=-0.05cm] (LLR13) {\footnotesize{LLR$(1,3)$}};
%\node[left of = n44,yshift=0.18cm,xshift=-0.05cm] (LLR14) {\footnotesize{LLR$(1,4)$}};
%\node[left of = n54,yshift=0.18cm,xshift=-0.05cm] (LLR15) {\footnotesize{LLR$(1,5)$}};

\node[left of = Pnode,node distance = 1.8cm] {Stage 1};
\node[left of = Pnode,node distance = 5.3cm] {Stage 2};
\node[left of = Pnode,node distance = 8.5cm] {Stage 3};
\end{tikzpicture}}
\caption{Tanner graph of a polar code of length $N_m=8$.}
\label{fig:polar_8}
\end{center}
\end{figure}
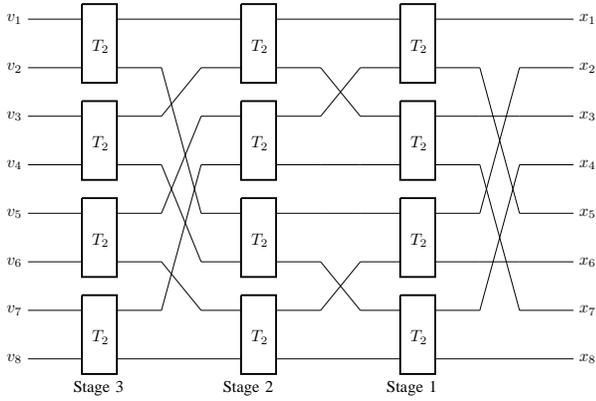

Originally, the polar code construction was designed for code lengths that are powers of two \cite{polar}; in the following we refer to those codes as \emph{mother polar codes}.
A mother polar code is an $(N_m,K_m)$ code of length $N_m=2^n$, for some integer $n$, and dimension $K_m$. 
A polar code is defined by a transformation matrix $T_{N_m} = T_2^{\otimes n}$, given by the $n$-fold Kronecker product of the kernel 
$ T_2 = 
\begin{bmatrix} 
        1 & 0 \\ 1 & 1 
\end{bmatrix}$, and a frozen set $\set{F} \subseteq \{1,\dots,N_m\}$ of size $N_m-K_m$. 
The frozen set is usually designed such that the error probability under SC decoding is minimized, i.e., the $N_m - K_m$ positions with the lowest bit reliabilities form the frozen set $\set{F}$.  
The reliabilities can be calculated by \emph{density evolution under a Gaussian approximation (DE/GA)}, tracking the mean value, variance, Batthacharyya parameter or mutual information of the Gaussian L-densities. 
An overview and comparison of such methods may be found in \cite{polar_const}. 
In the following we call \textit{information set} the complementary set of the frozen set, i.e., the set $\set{I} = \set{F}^c$ of size $K_m$. 
For illustration purposes, the Tanner graph of a polar code of length $N_m = 8$ is shown in Figure~\ref{fig:polar_8}. 

\subsection{Encoding of Polar Codes}
An information word $u \in \Fb^{K_m}$ is encoded into the codeword $x \in \Fb^{N_m}$ by defining a vector $v \in \Fb^{N_m}$ such that $v_\set{I} = u$ and $v_\set{F} = 0$, where $v_\set{I}$ and $v_\set{F}$ are the sub-vectors of $v$ defined by $\set{I}$ and $\set{F}$ respectively, and by computing the codeword as $x = v \cdot T_{N_m}$. 
In Arikan's paper \cite{polar}, the encoding was defined based on $\mathbf{B}_{N_m} \cdot T_{N_m}$ rather than $T_{N_m}$, where $\mathbf{B}_{N_m}$ denotes the bit-reversal permutation of length $N_m$. 
We recall that the bit-reversal permutation $\mathbf{B}_{N_m}$ is defined by indexing the sequence of numbers from $0$ to $N_m-1$ and then reversing the binary representations of each of them. 
We should note that the two codes obtained from these definitions are equivalent and differ only in the ordering of the code bits. 
We believe the description based on $T_{N_m}$ to be more convenient, and therefore we choose to use it in the following. 
For some constructions, the information word $u$ comprises a \emph{cyclic redundancy check}. 
If the CRC has length $N_\CRC$, the actual information word $u'$ has length $K'_m=K_m-N_\CRC$. 
The vector $u$, as defined above, is then formed by appending the vector $u_\CRC$ with the CRC bits as $u = \begin{bmatrix} u' & u_\CRC \end{bmatrix}$.

\subsection{Decoding of Polar Codes}
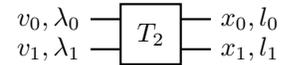
\begin{figure}[htb]
\begin{center}
	\begin{tikzpicture}[scale=0.2, line width=0.8pt]
	  \draw   (0,0) rectangle (4,4) node[midway]{$T_2$};
	  \draw[] (-2,3) node[left] {$v_0 , \lambda_0$} -- (0,3) ;
	  \draw[] (-2,1) node[left] {$v_1 , \lambda_1$} -- (0,1) ;
	  \draw[] (4,3)                     -- (6,3) node[right] {$x_0 , l_0$} ;
	  \draw[] (4,1)                     -- (6,1) node[right] {$x_1 , l_1$} ;
	\end{tikzpicture}
\caption{Basic decoder building block.}
\label{fig:2x2-block}
\end{center}
\end{figure}

Polar codes were originally designed to be decoded by the successive cancellation (SC) decoder proposed in \cite{polar}. 
The SC decoder is a very simple and fast decoder, however it may suffer from the propagation of decoding errors during the decoding process. 
The decoder was then extended to the successive cancellation list (SCL) decoder in \cite{list_decoding}, improving the BLER performance at the cost of an higher complexity. 
A SC decoder operates on the Tanner graph of the polar code, passing log-likelihood ratios (LLRs) from right to left and passing hard decisions on bits, also called partial sums, from left to right. 
The $2 \times 2$ building block, corresponding to $T_2$, of the SC decoder is depicted in Figure \ref{fig:2x2-block}, where $\lambda_i$ and $l_i$ denote the LLRs, while $v_i$ and $x_i$ denote the hard decisions.
For the kernel $T_2$, the hard-decision update rules are given by $x_0 = v_0 \oplus v_1$ and $x_1 = v_1$. 
The corresponding LLR update equations are $\lambda_0 =  l_0 \boxplus l_1$ and $\lambda_1 = (-1)^{u_0} \cdot l_0 + l_1$, where $ a \boxplus b \triangleq 2 \tanh^{-1} \bigl( \tanh\frac{a}{2} \cdot \tanh\frac{a}{2} \bigr) \approx  \sgn(a) \cdot \sgn(b) \cdot \min\{ |a|,|b| \}$. 
Considering the left end of the Tanner graph, when the LLR $\lambda_i$ on bit $u_i$ is known, the estimate on this bit is obtained as
\begin{equation}
  \hat{v}_i(\lambda_i) = 
    \begin{cases}
      0             &  \text{for $i \in \set{F}$,}                     \\
      0             &  \text{for $i \in \set{I}$ and $\lambda_i \geq 0$,} \\
      1             &  \text{for $i \in \set{I}$ and $\lambda_i < 0$.} \\
    \end{cases}
\end{equation}
The estimate on the information word is eventually $\hat{u} = \hat{v}_\set{I}$. 

For the SCL decoder, a list of $L$ most likely paths is stored instead of directly making hard decisions on $v_i$. 
Each path is identified by a vector $\hat{v}(l)$ denoting the sequence of hard decisions of this path and has a non-negative \emph{path metric} $\Lambda(l)$. 
This path metric is the sum of absolute values of the LLRs corresponding to those bits $\hat{v}(l)$ that are not the hard decisions of their LLRs, i.e., not following the hard decision of the LLR $\lambda_i$ yields a penalty on the path metric \cite{list_decoding}. 
For every path, a SC decoder is launched. 
Whenever the LLR $\lambda_i$ of a bit $v_i$ is calculated, each path in the list is extended by a zero and a one, creating two new extended paths, and the path metric for each candidate is updated. 
From the resulting $2L$ paths, the $L$ paths with the highest path metrics are deleted.  
When decoding is terminated, the path with the lowest path metric is chosen as the decoder output, and the information bits are extracted as for the SC decoder. 
In the case of a CRC-aided SCL decoder, the path with the lowest path metric that fulfills the CRC is chosen as the decoder output; if no path in the list fulfills the CRC, a decoding failure is declared. 
This decoder is referred to as \emph{CRC-aided SCL (CA-SCL) decoding} \cite{list_decoding}.

\section{Puncturing of Polar Codes}
\label{sec:punct}
In this section, we give an overview on puncturing of polar codes.

Puncturing is a technique used to obtain an $(N_m-P,K_m)$ code from an $(N_m,K_m)$ mother code for any $P<N_m-K_m$, thus increasing the code rate.  
As a result, $P$ code bits are not transmitted. 
The decoder handles punctured bits as erased, and applies the decoder of the mother code. 

In the case of polar codes, an $(N,K)$ punctured polar code is generated by a mother polar code of length $N_m = 2^{\lceil \log_2 N \rceil}$ and dimension $K_m = K$. 
The mother polar code is given by the transformation matrix $T_{N_m}$ and a frozen set $\set{F}$ of size $N_m - K$. 
The information word $u \in \Fb^{K}$ is encoded through the mother polar code into the codeword $x \in \Fb^{N_m}$. 
However, the vector $x$ contains $N_m > N$ bits, hence some of its bits will not be transmitted. 
If we call puncturing set $\set{P}$ the set of bits of $x$ not to be transmitted, the encoder transmits the vector $x_{\set{P}^C} \in \Fb^{N}$. 
At the decoder side, the channel LLRs for the punctured code bits are set to zero, and the decoder of the mother code is applied.

\begin{figure*}[!t]
\begin{center}
\subfloat[Decoding of the proposed $(6,4)$ punctured polar code.]{\label{fig:punct_dec} %\hspace{-0.75cm}
\resizebox{0.48\textwidth}{!}{ 
\includegraphics{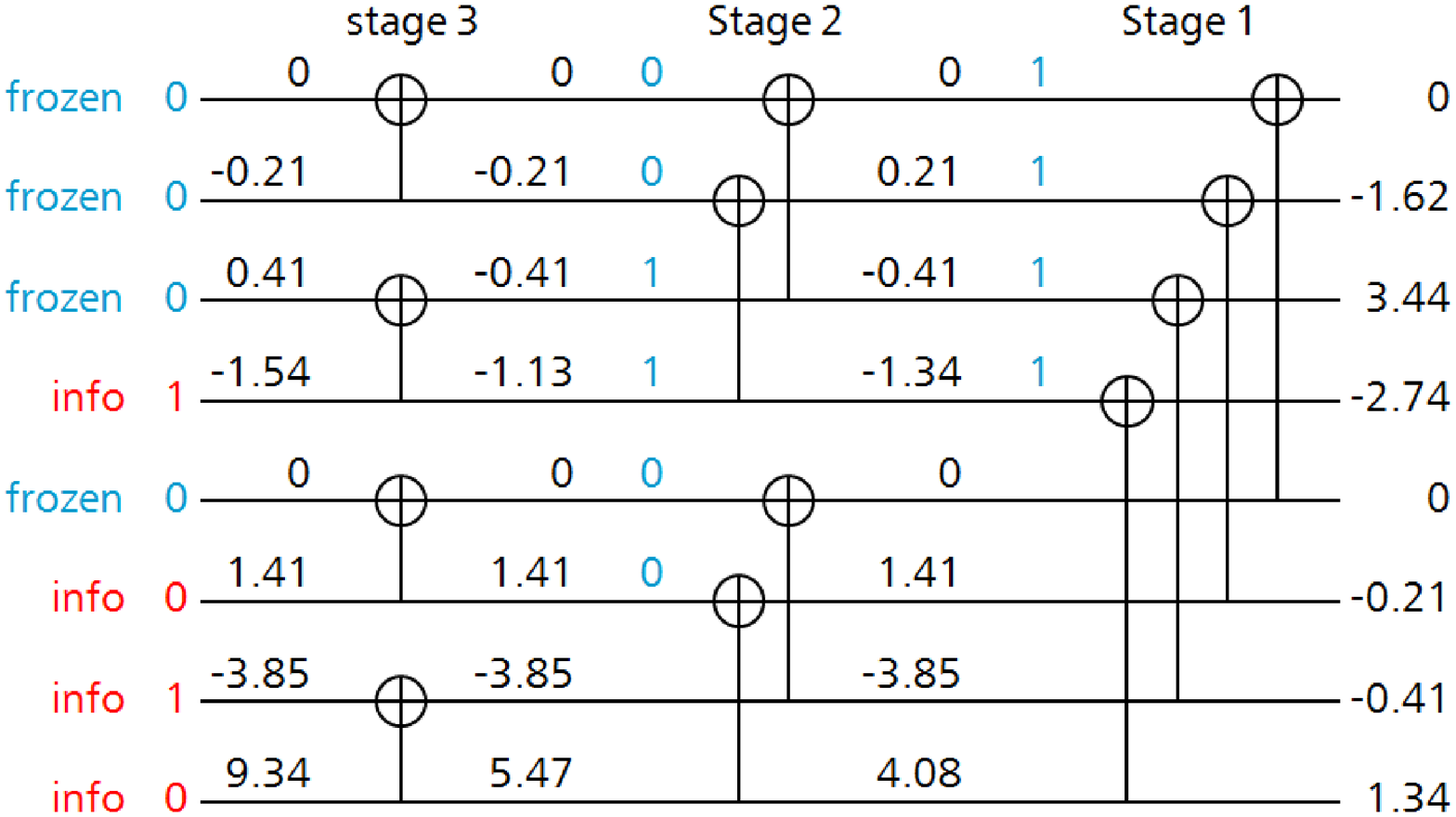}%png}
}
}
\subfloat[Decoding of the proposed $(6,4)$ shortened polar code.]{\label{fig:short_dec} %\hspace{-0.75cm}
\resizebox{0.48\textwidth}{!}{
\includegraphics{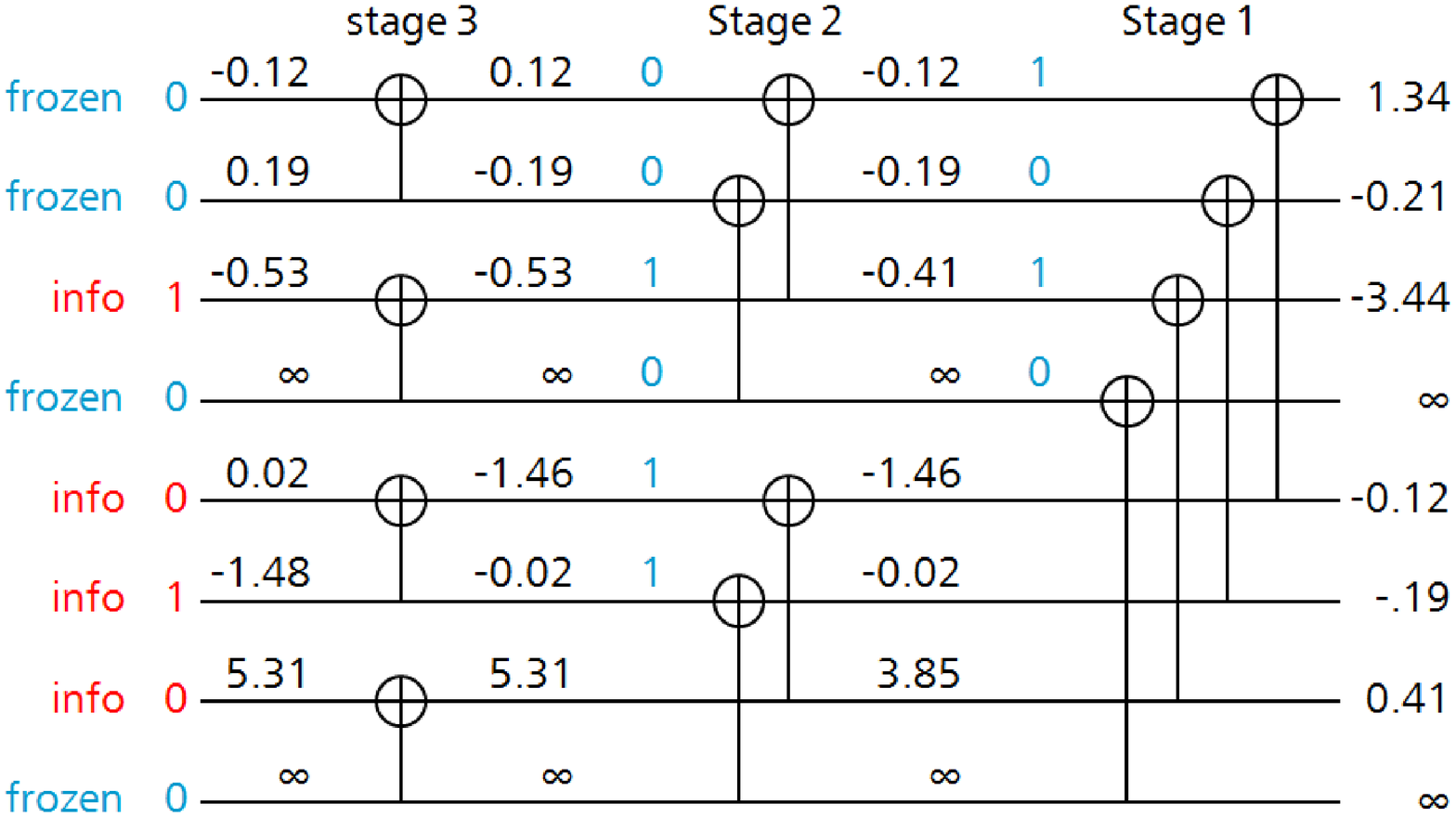}%png}
}
}
\caption{Polar decoding of a modified polar codes with $N_m = 8$.  Intermediate LLRs and partial sums are indicated. }
\label{fig:LLR_dec}
\end{center}
\vspace*{-0.45cm}
\end{figure*}

Frozen and puncturing sets are deeply connected. 
It is worth to notice that, while the frozen set $\set{F}$ is a subset of $v$, the puncturing set $\set{P}$ is a subset of the codeword $x$. 
In fact, the absolute value of the LLR is indicating the reliability of this bit. 
For punctured bits, the reliability is zero, hence these bits are completely unreliable for the decoder. 
When the SC decoder encounters such an unreliable position, its value cannot be resolved, and a LLR of value $\lambda_i = 0$ is injected into the decoder. 
The evolution of the unreliable positions in the decoding of a mother polar code of length $N_m = 8$ can be followed in Figure~\ref{fig:punct_dec}, where $\set{P} = \{1,5\}$. 
This unreliability is eventually projected to $v$, creating \emph{incapable bits}, i.e., bits for which it is not possible to take a non-random hard decision since they have a LLR of zero. 
This phenomenon is studied in \cite{short_mat}, where it is proved that the number of incapable bits is equal to the number of punctured bits. 
The set $\set{U}^{(\set{P})}$ of the incapable bits created by the puncturing set $\set{P}$, depicted as \emph{incapable set}, can be found by DE/GA or using the algorithm proposed in \cite{short_mat}. 
In the example of Figure~\ref{fig:punct_dec}, the incapable set is given by $\set{U}^{(\set{P})} = \{1,5\}$. 
As the value of such positions cannot be resolved by SC decoding, all incapable bits must be frozen bits to avoid an error floor. 
This gives the condition $\set{U}^{(\set{P})} \subseteq \set{F}$ in the design of the punctured polar code.
We should note that calculating the frozen set through the DE/GA algorithm from the punctured set, as in \cite{short_mat,chen_kai_punc}, naturally gives $\set{U}^{(\set{P})} \subseteq \set{F}$. 

\section{Shortening of Polar Codes}
\label{sec:short}
In this section, we give an overview on shortening of polar codes and we make some key observations for our proposed construction.

For shortening, a sub-code of the mother code is used in which $S$ bits of the codeword $x$ are restricted to a fixed value, e.g., zero. 
Shortening can be used to obtain an $(N_m-S,K_m-S)$ code from an $(N_m,K_m)$ systematic mother code for any $S<K$. 
In this case, $S$ information bits are chosen such that $S$ code bits are zero; these fixed code bits are not transmitted.  These code bits are referred to as shortened bits, and their index set is called the shortening set $\mathcal{S}$. 
At the decoder side, the shortened bits are known to be zeros, hence their LLRs are set to infinity (or in practice to sufficiently large values), and the decoder of the mother code is applied. 

Even if polar codes admit systematic construction \cite{syst_polar}, non-systematic polar codes can be shortened using a particular technique. 
In the general case, it can be complicated to determine $S$ bits of $v$ to be set to zero such that the $S$ code bits $x_{\set{S}}$ from the shortening set $\set{S}$ become zero, and thus are independent of the information bits.
In the case of polar codes, the constraints given by the presence of frozen bits naturally give bits that may be independent of the information bits.
In fact, if $S$ code bits are linear combinations of frozen bits only, they can be shortened since they are all zeros.
This translates into the constraint $T_n(\set{I},\set{S}) = \mathbf{0}$, where $T_n(\set{I},\set{S})$ represents the sub-matrix of $T_n$ formed by the rows in $\set{I}$ and the columns in $\set{S}$. 
With an abuse of notation\footnote{Since $\set{S}$ deals with the codeword $x$ and $\set{F}$ deals with $v$, they lie in different vectors.}, if $\set{S} \subseteq \set{F}$, i.e., the shortened bits are frozen, we have that $T_n(\set{S}^C,\set{S}) = \mathbf{0}$ implies $T_n(\set{I},\set{S}) = \mathbf{0}$.
 
In the case of non-systematic mother polar codes, an $(N,K)$ shortened polar code is generated by a mother polar code of length $N_m = 2^{\lceil \log_2 N \rceil}$ and dimension $K_m = K$, similarly to punctured polar codes. 
Frozen and shortening set have to be selected such that $T_n(\set{I},\set{S}) = \mathbf{0}$. 
The information word $u \in \Fb^{K}$ is encoded through the mother polar code into the codeword $x \in \Fb^{N_m}$, and the vector $x_{\set{S}^C} \in \Fb^{N}$ is transmitted, while the bits $x_{\set{S}}$ are shortened. 
At the decoder side, the channel LLRs for the shortened code bits are initialized with sufficiently large positive values (in theory with infinity), and the decoder of the mother code is applied.

We have seen that non-systematic shortened polar codes have a construction which is similar to the one of punctured polar codes. 
Due to this reason, there has been some confusion in the literature on the nature of shortened polar codes, in particular shortening was sometimes referred to as puncturing, e.g. in \cite{wang_liu}. 
We recall here that a code is shortened when code bits are forced to take fixed values (typically zero) by appropriate encoding and then eliminated, while a code is punctured when code bits with non-fixed values are eliminated. 
As a result, while both shortening and puncturing reduce the length of the code, shortening may also reduce the dimension of the code. 
However, polar codes can be shortened without reducing the code dimension. 

The introduction of code bits of $x$ that are known at the decoder creates some high reliability bits in $v$, that we call \emph{overcapable bits}. 
The evolution of the high reliability positions in the decoding can be followed in Figure~\ref{fig:short_dec}, where $\set{S} = \{4,8\}$. 
In this case, the overcapable set, i.e., the set of overcapable bits generated by $\set{S}$, is given by $\set{O}^{(\set{S})} = \{4,8\}$. 
An overcapable bit $v_i$, $i \in \set{O}^{(\set{S})}$, is surely correctly decoded by a SC decoder, i.e., the probability to commit an error decoding that bit is zero. 
It can be proved that, similarly to incapable bits, the number of overcapable bits is equal to the number of shortened bits. 
In theory, the presence of high reliability bits should increase the BLER performance of the code, but $T_n(\set{I},\set{S}) = 0$ implies that $\set{O}^{(\set{S})} \subseteq \set{F}$, i.e., in a non-systematic shortened polar code the overcapable bits are all frozen. 
However, even if in a systematic shortened polar code the overcapable set can be designed to be a subset of the information set, the augmented dimension of the code makes this design less appealing in terms of BLER performance.

\section{Proposed Framework}
\label{sec:prop}
In this section, we describe our low-complexity unified framework to modify polar codes. 
In our proposal, the frozen set is selected according to the bit reliabilities of the mother polar code, while puncturing and shortening sets are generated based on the bit-reversal permutation \cite{polar}, denoted by the index vector $\mathbf{B}_{N_m}$ of length $N_m$. 
For sake of clearness, we will give an example of encoding and decoding of modified polar codes based on the proposed construction. 
In particular, we describe the encoding and decoding of $(6,4)$-codes based on puncturing and shortening of a mother polar codes. 
In this case, the mother polar code has length $N_m = 2^{\lceil \log_2 6 \rceil} = 8$ and dimension of $K_m = 4$. 
The reliability permutation of the mother polar code is given by $\mathbf{R}_{N_m} = (1,2,3,5,4,6,7,8)$, ordered in increasing reliability, and the bit-reversal permutation is $\mathbf{B}_{N_m} = (1,5,3,7,2,6,4,8)$. 
For both codes, we fix the information word as $u = [1,0,1,0]$. 
For illustration, we depict the complete procedure in Figure~\ref{fig:LLR_dec}. 

\subsection{Puncturing}
An $(N,K)$ \emph{punctured polar code} is generated from an $(N_m,K_m)$ mother polar code with $K_m = K$ and $N_m = 2^{\lceil \log_2 N \rceil}$. 
The puncturing set is given by $\set{P} = \mathbf{B}_{N_m}(1,\dots,P)$ with $P = N_m - N$, where $\mathbf{B}_{N_m}(1,\dots,P)$ denotes the entries of $\mathbf{B}_{N_m}$ with indices $1,\dots,P$, i.e., the first $P$ bits of $x$ chosen in bit-reversal order are punctured. 
With this construction, we obtain $\set{U}^{(\set{P})} = \set{P}$, i.e., the bits of $v$ corresponding to the positions in $\set{P}$ are incapable bits; therefore they form a subset of the frozen set. 
To design the remaining frozen bits, we use the reliabilities of the mother polar code. 
If $\mathbf{R}_{N_m}$ is the permutation listing the bits of $v$ in reliability order from the least reliable to the most reliable, the frozen set $\set{F}$ is given by the union of $\set{P}$ and the first $N-K$ elements of $\mathbf{R}_{N_m} \setminus \set{P}$. 
The remaining $K$ bits of $v$ are set as information bits.   
%Using $\mathcal{F}$ and $\mathcal{P}$, encoding and decoding is performed as described before. 

In the case of the $(6,4)$ \emph{punctured code} based on the previously described mother code, the puncturing set is given by $\set{P} = \mathbf{B}_{N_m}(1,2) = \{ 1,5 \}$, since $P = 2$. 
Due to $\set{U}^{(\set{P})} = \set{P}$, the corresponding input bits have to be frozen. 
The remaining frozen bits are the first $N-K = 2$ elements of $\mathbf{R}_{N_m} \setminus \set{P} = \{ 2,3,4,6,7,8 \}$, i.e., the frozen set of the mother polar code is given by $\set{F} = \{ 1,2,3,5 \}$. 
As a consequence, $v = [0,0,0,1,0,0,1,0]$ and the codeword of the mother polar code is given as $x = [0,1,0,1,1,0,1,0]$. 
After puncturing, the vector $x_{\set{P}^c} = [1,0,1,0,1,0]$ is transmitted. 
After reception of the channel outputs, the decoder calculates the LLRs of the received symbols. These LLRs are provided to the decoder of the mother polar code, while the LLRs of punctured code bits are set to zero. 
The decoding procedure is described in Figure~\ref{fig:punct_dec}. 

\subsection{Shortening}
The generation of a \emph{shortened polar code} is done in a similar fashion. 
A $(N,K)$ shortened polar code is generated from a $(N_m,K_m)$ mother polar code with $K_m = K$ and $N_m = 2^{\lceil \log_2 N \rceil}$. 
The shortening set is given by $\set{S} = \mathbf{B}_{N_m}(N_m-S+1,\dots,N_m)$ with $S = N_m - N$, i.e., the last $S$ bits of $x$ chosen in bit reversal order are shortened. 
With this construction, we obtain $\set{O}^{(\set{S})} = \set{S}$, i.e., the bits of $v$ corresponding to the positions in $\set{S}$ are overcapable bits; therefore they form a subset of the frozen set.  Notice the similarity to puncturing above.
The remaining frozen bits are chosen based on the reliability order of the mother polar code. 
As before, the frozen set $\set{F}$ is given by the union of $\set{S}$ and the first $N-K$ elements of $\mathbf{R}_{N_m} \setminus \set{S}$. 
The remaining $K$ bits of $v$ are set as information bits. 
%Using $\mathcal{F}$ and $\mathcal{S}$, encoding and decoding is performed as described before. 

For the $(6,4)$ \emph{shortened code}, the shortening set is given by $\set{S} = \mathbf{B}_{N_m}(7,8) = \{ 4,8 \}$, since $S = 2$, and these positions are added to the frozen set due to $\set{O}^{(\set{S})} = \set{S}$. 
The remaining frozen bits are the first $N-K = 2$ elements of $\mathbf{R}_{N_m} \setminus \set{S} = \{ 1,2,3,5,6,7 \}$, i.e., the frozen set of the mother polar code is given by $\set{F} = \{ 1,2,4,8 \}$. 
As a consequence, $v = [0,0,1,0,0,1,0,0]$ and the codeword of the mother polar code is $x = [0,1,1,0,1,1,0,0]$. 
After shortening, the vector $x_{\set{S}^c} = [0,1,1,1,1,0]$ is transmitted. 
After reception of the channel outputs, the decoder calculates the LLRs of the received symbols.  These LLRs are provided to the decoder of the mother polar code, while the LLRs of the shortened code bits are set to $+\infty$. 
The decoding procedure is described in Figure~\ref{fig:short_dec}.

\subsection{Complexity Discussion}
We propose bit-reversal permutation, necessary in the construction of polar codes, as a basis for puncturing and shortening sets due to the regularity it transmits to incapable and overcapable sets, for which $\set{U}^{(\set{P})} = \set{P}$ and $\set{O}^{(\set{S})} = \set{S}$ respectively. 
The knowledge of the incapable and overcapable sets is of paramount importance to the BLER performance of the modified polar codes, in particular for extreme rates. 
If the reliability permutation of the mother polar code is stored, the code design is performed with a complexity of $O(K)$ by selecting the last $K$ bits of $\mathbf{R}_{N_m}$ not belonging to $\set{P}$ or $\set{S}$.

On the other hand, in DE/GA-based frameworks like \cite{wang_liu,chen_kai_punc}, the reliability of the bits has to be calculated on-the-fly, at a complexity of $O(N \log N)$. 
The algorithm proposed in \cite{isit_punc} to retrieve the incapable set has comparable complexity, but its implementation does not require the evaluation of a non-linear function for reliabilities calculation \cite{polar_const}.

\section{Numerical Illustrations}
\label{sec:num}
\begin{figure}
\includegraphics[width=0.48\textwidth]{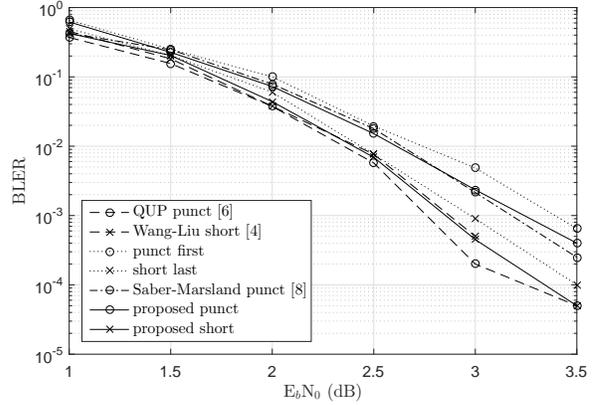}
\caption{Performance of modified $(320,160)$ polar codes (rate $0.5$) under CA-SCL decoding, with list size $L=32$ and 24 CRC bits.}
\label{fig:plot_320_160}
\end{figure}

\begin{figure}
\includegraphics[width=0.48\textwidth]{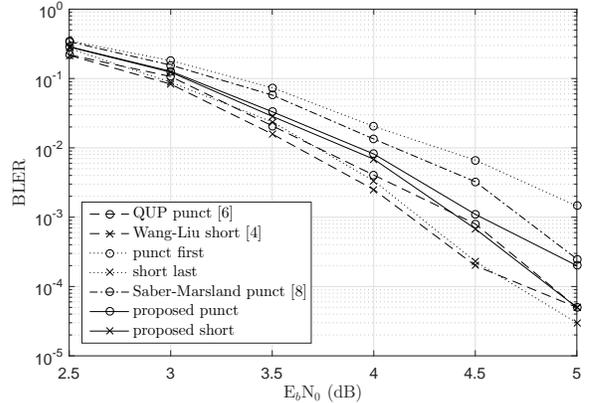}
\caption{Performance of modified $(160,120)$ polar codes (rate $0.75$) under CA-SCL decoding, with list size $L=32$ and 24 CRC bits.}
\label{fig:plot_160_120}
\end{figure}

\begin{figure}
\includegraphics[width=0.48\textwidth]{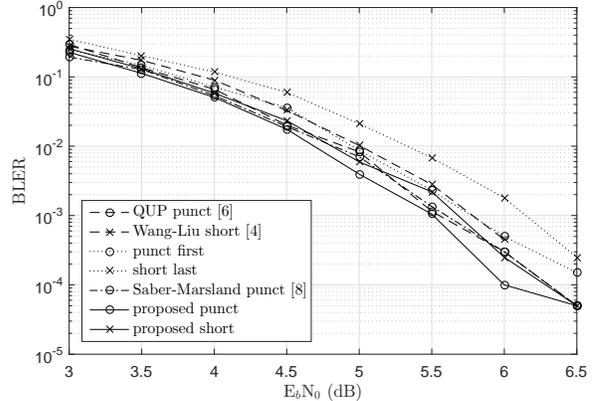}
\caption{Performance of modified $(160,40)$ polar codes (rate $0.25$) under CA-SCL decoding, with list size $L=32$ and 24 CRC bits.}
\label{fig:plot_160_40}
\end{figure}

In this section, we show the performance of the proposed puncturing and shortening schemes in terms of BLER. 
In Figures~\ref{fig:plot_320_160}-\ref{fig:plot_160_40} we compare our modified polar codes to the state-of-the-art of puncturing and shortening schemes of polar codes for transmission over an additive white Gaussian noise (AWGN) channel with QPSK. 
All the evaluated codes are decoded using CRC-aided SC list decoding with $24$ CRC bits, as for the current turbo coding scheme in 3GPP-LTE,  and list size $L=32$.
We compare our solution to various other puncturing and shortening techniques of different complexity. 

We observe that the technique giving the best results in terms of BLER consists in first designing the puncturing or shortening set, and then calculating the reliability permutation of the resulting code. 
In practice, DE/GA (or an alternative algorithm) is run to calculate the mean of the L-density of each bit of $v$, where the means of the punctured bits are set to zero, the means of the shortened bits are set to a sufficiently large value (theoretically infinity), and the means of the other code bit positions are set to the value corresponding to the SNR of the AWGN channel.
In \cite{chen_kai_punc}, the authors use this technique to calculate the frozen set of a polar code where, according to our definition of the transformation matrix, the first $P$ bits of $x$ are punctured. 
Similarly, in \cite{wang_liu} the authors calculate the frozen set of a polar code where the last $S$ bits of $x$ are shortened (and the last $S$ bits of $v$ are frozen). 
These codes are used here as a performance reference, since they provide very good BLER performance, albeit at the cost of a higher computational complexity: the frozen set calculated with this technique does not have a structure that is easy to describe, and hence has to be calculated or stored for every code length and rate of interest, increasing the latency of polar encoders and decoders. 
This added complexity makes this technique unsuitable for practical scenarios, where larger ranges of lengths and rates need to be covered.

In order to decrease the overall complexity, the reliability permutation of the mother polar code can be calculated offline and stored, while puncturing and shortening sets are calculated on-the-fly based on that.
This idea is used for example in \cite{ARQ_punct}, where the code bits in $x$ are punctured following the reliability order $\mathbf{R}_{N_m}$ of $v$. 
However, puncturing in reliability order does not guarantee that $\set{U}^{(\set{P})} \subseteq \set{F}$, making this technique less suitable for high code rates, as shown in Figure~\ref{fig:plot_160_120}.  
Similarly, it is possible to puncture the first $P$ bits of $x$, as proposed in \cite{chen_kai_punc}, but without performing DE/GA, in order to evaluate the effect of the puncturing set on the reliability order. 
This results in making the first $P$ bits of $v$ incapable, i.e., $\set{U}^{(\set{P})} = \set{P}$, with the same shortcoming presented in \cite{ARQ_punct} for high code rates, as shown in Figure~\ref{fig:plot_160_120}.   
For shortening, it is possible to shorten the last $S$ bits of $x$, as proposed in \cite{wang_liu}, without performing DE/GA, again to evaluate the effect of the shortening set on the reliability order.  
However, this method results in freezing the last bits of $v$, which are usually the most reliable bits, hence resulting in a worst BLER performance for low-rate codes, as shown in Figure~\ref{fig:plot_160_40}. 

In order to mitigate the effect of the puncturing and shortening set on the reliability order of the mother polar code in the low-complexity design of modified polar codes, we propose to calculate the puncturing and shortening sets on-the-fly based on the bit-reversal permutation. 
For puncturing, our solution assures $\set{U}^{(\set{P})} \subseteq \set{F}$, which is advantageous compared to other low-complexity puncturing strategies which do not possess this feature. 
In Figures~\ref{fig:plot_320_160}-\ref{fig:plot_160_40} we can see that the proposed puncturing shows good BLER performance compared to the other low-complexity scheme, without suffering from the high-rate error floor, as shown in Figure~\ref{fig:plot_160_120}. 
For shortening, our proposal allows one to pick frozen bits among the less reliable bits as well, not focusing on the most reliable bits only. 
This mitigates the BLER performance problem for low-rate codes, as shown in Figure~\ref{fig:plot_160_40}. 

\section{Conclusions}
\label{sec:conclusions}
In this work, we reviewed shortening and puncturing techniques for polar codes, and proposed a simple new solution with pre-computed permutations to determine the frozen sets and puncturing and shortening pattern based on the bit-reversal permutation. 
While several independent puncturing and shortening designs were attempted in the literature, we aimed at a unifying construction encompassing both techniques in order to achieve any code length and rate at low complexity. 
Our solution significantly reduces the complexity as compared to state-of-the-art proposals while maintaining a block error rate performance comparable to these highly optimized constructions.  
Hence our proposed design with a compact and low-complexity description of the constructed polar codes allows for practical application in future communication systems where a large set of polar codes with different lengths and rates are required. 
%Future work will address the joint application of puncturing and shortening to further improve the error rate performance.
Future work will address the joint application of puncturing and shortening to further decrease the error rate.

\bibliographystyle{IEEEbib}
\bibliography{polar_codes_bib}

\end{document}